\documentclass[conference]{IEEEtran}
\IEEEoverridecommandlockouts

\usepackage{cite}
\usepackage{amsmath,amssymb,amsfonts}
\usepackage{algorithm}
\usepackage{orcidlink}
\usepackage{algpseudocode} 
\usepackage{graphicx}
\usepackage{textcomp}
\usepackage{xcolor}
\usepackage{multirow}
\usepackage{booktabs}
\usepackage{footnote}
\usepackage{caption, subcaption}
\usepackage{float}
\usepackage{url}
\usepackage{hyperref}
\makeatletter
\newcommand{\etal}{\textit{et al.}}
\newcommand{\eg}{\textit{e.g.}}
\newcommand{\ie}{\textit{i.e.}}

\newcommand*{\rom}[1]{\expandafter\@slowromancap\romannumeral #1@}
\makeatother

\newcommand{\lili}[1]{{\color{blue}{\bf }{\bf lili: }#1}}   
\newcommand{\kj}[1]{{\color{green}{\bf KJ: }#1}} 
\newcommand{\zhiyuan}[1]{{\color{purple}{\bf Zhiyuan: }#1}} 
 
\def\BibTeX{{\rm B\kern-.05em{\sc i\kern-.025em b}\kern-.08em
    T\kern-.1667em\lower.7ex\hbox{E}\kern-.125emX}}

\renewcommand{\kj}{}
\renewcommand{\zhiyuan}{}
\renewcommand{\lili}{}

\graphicspath{{figures/}}

\long\def\comment#1{}
\long\def\delete#1{}
\newcommand{\para}[1]{\smallskip\noindent {\bf #1}}

\newcommand{\tsc}[1]{\textsuperscript{#1}}

\title{Joint Optimization of Handoff and Video Rate in LEO Satellite Networks}

\makeatletter
\renewcommand{\footnoterule}{
    \kern -3pt
    \hrule width 0.4\columnwidth
    \kern 2.6pt
}
\makeatother

\author{
\IEEEauthorblockN{\small
Kyoungjun Park\orcidlink{0000-0002-5028-2697}\IEEEauthorrefmark{2}, Zhiyuan He\orcidlink{0009-0004-6891-6119}\IEEEauthorrefmark{3}\tsc{,1}, Cheng Luo\orcidlink{0009-0006-9505-5830}\IEEEauthorrefmark{2}\tsc{,1,2}, Yi Xu\orcidlink{0009-0002-1040-2110}\IEEEauthorrefmark{4}, Lili Qiu\orcidlink{0009-0003-8131-7439}\IEEEauthorrefmark{2}\IEEEauthorrefmark{3}\thanks{Corresponding author: Lili Qiu, liliqiu@cs.utexas.edu}
Changhan Ge\orcidlink{0000-0002-0873-8179}\IEEEauthorrefmark{2}\tsc{,3}, Muhammad Muaz\orcidlink{0009-0000-7922-6456}\IEEEauthorrefmark{2}
}
\IEEEauthorblockA{\IEEEauthorrefmark{2}\textit{The University of Texas at Austin}, Austin, Texas, United States of America, \{kjpark, lili, chge, mmuaz\}@cs.utexas.edu\hspace{0.25cm}\\
\IEEEauthorrefmark{3}\textit{Microsoft Research Asia}, Shanghai, People's Republic of China, \{zhiyuhe, luocheng, liliqiu\}@microsoft.com\\
\IEEEauthorrefmark{4}\textit{University of Science and Technology of China}, Hefei, Anhui, People's Republic of China, yi\_xu@mail.ustc.edu.cn}
}

\begin{document}

\setlength{\abovecaptionskip}{2pt}
\setlength{\belowcaptionskip}{2pt}
\setlength{\abovedisplayshortskip}{0pt}
\setlength{\belowdisplayshortskip}{0pt}
\setlength{\textfloatsep}{0pt}

\maketitle

\begingroup
\renewcommand\thefootnote{}\footnotetext{\tsc{1} These authors contributed equally.}
\renewcommand\thefootnote{}\footnotetext{\tsc{2} Cheng Luo is currently a post-doctoral researcher at California Institute of Technology, Pasadena, CA, USA}
\renewcommand\thefootnote{}\footnotetext{\tsc{3} Changhan Ge is currently a research engineer at AT\&T Labs - Research, Bedminster, NJ, USA.}
\renewcommand\thefootnote{}\footnotetext{\tsc{4} \href{https://github.com/kyoungjunpark/Joint-LEO}{https://github.com/kyoungjunpark/Joint-LEO}}
\addtocounter{footnote}{0}
\endgroup

\begin{abstract}
Low Earth Orbit (LEO) satellite communication is a promising approach to providing Internet connectivity to users in many remote areas. As videos are likely to account for most traffic in the LEO satellite network, as in the rest of the Internet, this work introduces a novel video-aware mobility management framework tailored for LEO satellite networks. Utilizing simulation models alongside real-world datasets, we show the importance of handoff strategy and throughput prediction algorithms in single-user and multi-user video streaming scenarios. Motivated by these observations, we propose a set of novel algorithms that can jointly choose the satellite and video bitrate to optimize the Quality of Experience (QoE). We first develop Model Predictive Control (MPC) and Reinforcement Learning (RL) based algorithms for a single user, and then extend them to accommodate multiple competing users that may share the same satellite. We introduce \textit{centralized training and distributed inference} for our RL design, enabling a distributed policy informed by a global perspective. We demonstrate the effectiveness of our proposed models using trace-driven simulation and testbed experiments.  We share our code and data with the research community\tsc{4}.
\end{abstract}



\section{Introduction}
\label{sec:introduction}

\para{Motivation:} Recently, we have witnessed a rapid increase in Low Earth Orbit (LEO) satellite networks. Several companies, such as SpaceX, Amazon, and OneWeb, have launched or planned to launch thousands of satellites into space to build their commercial satellite networks. LEO satellite networks are attractive due to their ability to provide global network coverage at an affordable price and their lower latency and higher throughput than geostationary satellites \cite{chitre1999next, jamalipour2001role}. On the other hand, a LEO satellite moves at a rapid speed -- it moves at around $7.5~km/s$ relative to ground stations, circulating the Earth every 90 to 110 minutes. This rapid movement restricts a satellite connection to a duration of only tens of seconds to three minutes, thus necessitating frequent handoffs. Zhao \etal~\cite{zhao2023realtime} report that Starlink switches the primary link to another satellite approximately once every 15 seconds. Therefore, such frequent handoffs call for an effective handoff strategy to ensure smooth and high performance. 

Note that although the handoff frequency in LEO satellite communication is high, the satellites' movement follows a deterministic pattern. This characteristic is very different from traditional Wi-Fi and cellular networks, where mobility is less predictable. Such a predictable mobility pattern presents a valuable opportunity to optimize handoff strategies, yielding significant potential benefits.

Meanwhile, we also observe a rapid growth in video streaming traffic on the Internet. Video streaming has become the majority of Internet traffic. 
LEO satellite providers offer Internet access to remote regions (e.g., Africa, remote villages), where satellite links are their only way to access the Internet. We expect remote users to have similar traffic patterns as traditional Internet users to enjoy video streaming, which accounts for most Internet traffic.


Studies have shown that video quality is crucial to retaining viewers \cite{dobrian2011understanding, park2022neusaver}. Adaptive video streaming addresses fluctuating network throughput, where the sender divides a video into several chunks and encodes each video chunk at several different data rates \cite{10631023}. Clients dynamically select the appropriate bitrate based on current network conditions. A variety of algorithms have been proposed to select the video bitrate to optimize the video Quality of Experience (QoE), which is determined by three major factors: video quality, rebuffering time, and smoothness of quality. Some use optimization while others use Reinforcement Learning (RL) to adapt the video bitrate. 

\para{Our approach:}  While video streaming and LEO network are important topics, little work studies them jointly. This motivates our exploration of this topic. We first conduct simulations and real-world measurements to identify key challenges and opportunities in the video streaming process in LEO networks. Our results indicate that (i) throughput pattern is important in the video QoE and we should explicitly incorporate the LEO satellites' movement along with the environmental factors for prediction, (ii) widely used handoff algorithms, such as maximizing Received Signal Strength (RSS) or maximizing the satellite serving time, do not yield good video QoE, and (iii) multiple users' sharing the satellite's bandwidth further complicates the design of video rate adaptation.

To address these challenges, we observe a strong interaction between handoff and Adaptive Bitrate (ABR) algorithms and propose novel methods that jointly optimize satellite selection and video bitrate. We show that by explicitly incorporating the impact of satellite selection into video QoE, we can effectively balance the tradeoff between handoff numbers and RSS. Moreover, to support multiple users competing for limited satellite throughput, we design novel algorithms to jointly select satellites and video bitrates for all users. 

Our contributions can be summarized as follows: 
\begin{itemize}
    \item \lili{We first use measurements to identify challenges of video streaming in LEO satellite networks. We show that it is important to design a handoff strategy considering video performance. We also show the importance of throughput prediction in both single-user and multi-user scenarios.} 
    
   \item We develop the first algorithms that jointly manage mobility and video bitrate in LEO networks. Our suggested algorithms are based on Model Predictive Control (MPC) and RL. In particular, we develop a \textit{centralized training and distributed inference-based RL algorithm} to explicitly consider the interaction between multiple satellites and ground stations while supporting practical use.
   
  \item We conduct both trace-driven simulation and testbed experiments. Our results show that joint video rate and satellite selection lead to 18\% to 68\% QoE improvement across different settings.

\end{itemize}

\section{Related Works}
\label{sec:related}

\noindent \textbf{LEO satellite communication.}
LEO satellites orbiting at $250$–$1000~\mathrm{km}$ offer lower delay than Geostationary Earth Orbit (GEO) and Medium Earth Orbit (MEO) systems due to their proximity to Earth. While electromagnetic signals propagate at light speed in space, optical fiber slows signals by approximately 31\%~\cite{latencyopticalfiber}. As a result, LEO constellations can achieve round-trip latency near 20~ms~\cite{handley2018delay}, enabling real-time applications such as video conferencing and cloud gaming. For long-distance communication (beyond 3000~km), LEO satellites can outperform terrestrial fiber links in latency~\cite{handley2018delay}. Their low altitude also supports the use of high-frequency bands (e.g., $K_u$, $K_a$, THz), which allow higher throughput.

Recent works focus on efficient networking and scheduling for LEO systems. Starfront~\cite{lai2021cooperatively}, Orbitcast~\cite{lai2021orbitcast}, and L2D2~\cite{vasisht2021l2d2} improve downlink performance using trajectory-based scheduling, satellite visibility estimation, and ground station coordination. SkyTube~\cite{lin2023space} jointly optimizes satellite selection and video resolution using local user states.

\noindent \textbf{Bitrate adaptation.}
ABR algorithms dynamically select video quality to maximize QoE under varying networks. Rate-based schemes, such as FESTIVE~\cite{jiang2012improving} and CS2P~\cite{sun2016cs2p}, rely on predicted throughput. Buffer-based approaches, e.g., BOLA~\cite{spiteri2020bola}, use buffer occupancy to guide decisions. Hybrid methods, including RobustMPC~\cite{yin2015control}, combine both metrics for better performance. RL has been applied to ABR via Pensieve~\cite{mao2017neural}, which uses the A3C algorithm~\cite{mnih2016asynchronous} to adapt to unseen network traces. QARC~\cite{huang2018qarc} enhances this by learning representations of video-aware states using CNNs and RNNs. Fugu~\cite{yan2020learning} replaces MPC throughput predictors with learned models. SENSEI~\cite{zhang2021sensei} incorporates human quality sensitivity into the adaptation logic. Other efforts explore multipath-aware ABR (PERM~\cite{guan2020perm}) and low-latency streaming (BitLat~\cite{wang2019bitlat}).

\noindent \textbf{Satellite handoff.}
Due to their high orbital velocity (7.5~km/s~\cite{9771894}), LEO satellites periodically leave the field of view of ground users, necessitating frequent handoffs. These transitions introduce 150–300~ms of disconnection at the physical layer~\cite{yang2016seamless}, with compounding effects on transport and application layers~\cite{li2017state}. Earlier studies addressed mobility in LEO systems from an IP- or circuit-switching perspective~\cite{mobility-LEO1, mobility-LEO2}. Recent work by~\cite{juan2022handover} proposes trajectory-aware handoff timing algorithms, simulating the impact of mobility and antenna tracking errors on performance.

\section{Background and Motivation}
\label{sec:motivate}

This section presents key challenges and opportunities in supporting video streaming over LEO satellite networks, motivating our algorithmic design.

\subsection{Predictability of Satellite Signal}
\label{sec:predictablility}

LEO satellites follow predictable orbits, allowing accurate signal forecasting. Although Starlink terminals lack RSS reporting, we measure the Signal-to-Noise Ratio (SNR) from NOAA satellites with similar motion and altitude. 

\begin{figure}[tp]
    \centering
    \begin{subfigure}{0.5\columnwidth}
        \centering
        \includegraphics[width=1\textwidth]{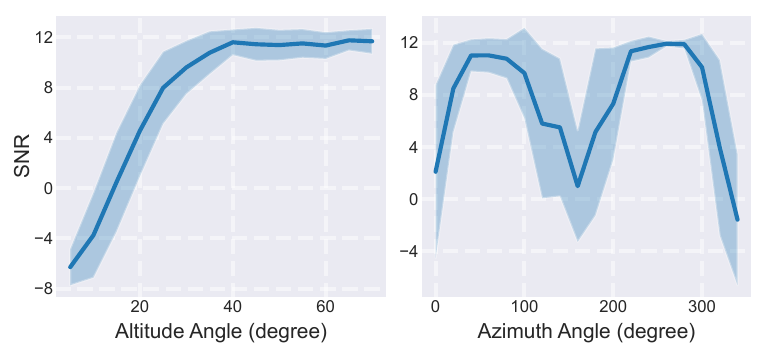}
        \caption{SNR vs. altitude/azimuth}
        \label{fig:noaa-snr-relationship}
    \end{subfigure}
    \begin{subfigure}{0.48\columnwidth}
        \centering
        \includegraphics[width=1\textwidth]{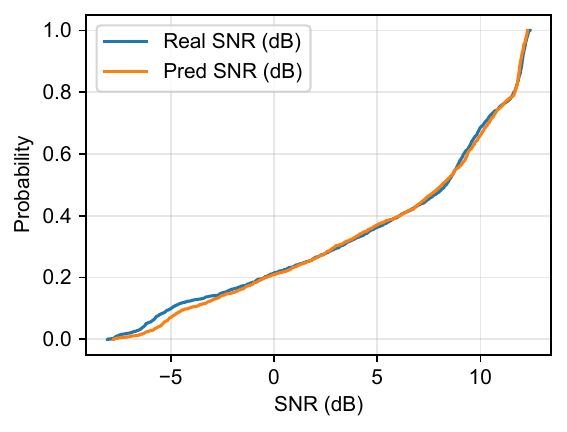}
        \caption{Predicted vs. measured SNR}
        \label{fig:snr-cdf}
    \end{subfigure}
    \caption{SNR characteristics of NOAA satellites.}
\end{figure}

Fig.~\ref{fig:noaa-snr-relationship} shows a strong correlation between SNR and satellite angles. We develop an ML-based SNR predictor achieving a Mean Absolute Error (MAE) of $0.84$~dB (Fig.~\ref{fig:snr-cdf}). This predictability enables ABR algorithms~\cite{jiang2012improving, sun2016cs2p, yin2015control} to map SNR to data rates and expected video bitrate, enhancing streaming performance.

\subsection{Video Performance in LEO Network}
\label{ssec:motivate-video}

\begin{figure}[!t]
    \centering
    \includegraphics[width=\columnwidth]{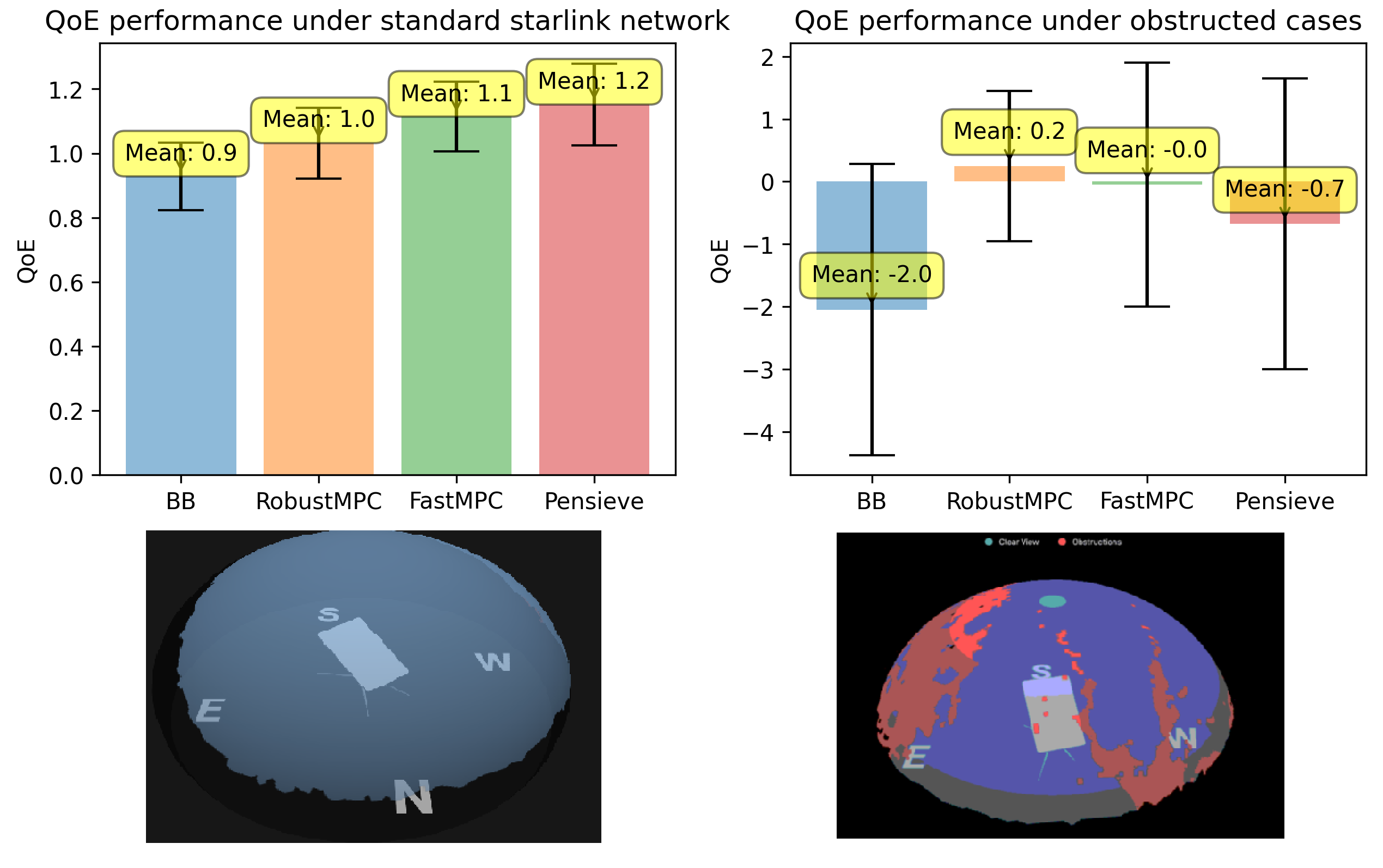}
    \caption{Video QoE under unobstructed (left) and obstructed (right) settings, with visibility maps from the Starlink app.}
    \label{fig:motivation_video_QoE_stanard}
\end{figure}

We study video QoE under real Starlink usage in two environments: one free of obstructions and another with visible obstructions (\eg, buildings or trees). As shown in Fig.~\ref{fig:motivation_video_QoE_stanard}, obstructions severely degrade QoE by causing throughput drops below $0.1$~Mbps. 

We use popular ABR algorithms such as RobustMPC~\cite{yin2015control} and Pensieve~\cite{mao2017neural}, and apply the following QoE metric~\cite{yin2015control}:
\begin{align}
\mathrm{QoE}
=
&\sum_{k=1}^{N} \left[\mu_{1} Q(R_{k}) - \mu_{2} T(R_{k})\right]\nonumber\\
&- \mu_{3} \sum_{k=1}^{N-1}
\left| Q(R_{k+1}) - Q(R_{k}) \right|
\label{eq:QoE}
\end{align}
where $R_k$ is the bitrate for chunk $k$, $Q()$ is the bitrate-to-quality mapping, $T()$ is the rebuffering time, and $\mu_i$ are weighting parameters, respectively.

By utilizing SNR predictions, ABR algorithms can proactively reduce the video bitrate prior to signal degradation, allowing buffer accumulation and enhancing robustness. Moreover, jointly switching to an alternative satellite before disconnection helps sustain consistent throughput and QoE.

\subsection{Satellite Handoff Selection}
\label{sec:motivate-handover}

Due to rapid satellite motion ($7.5$~km/s~\cite{9771894}), frequent handoffs are inevitable. Each handoff incurs $150$--$300$~ms delay~\cite{yang2016seamless}, amplified at higher layers~\cite{li2017state}, sometimes leading to 5-second buffering~\cite{zhao2023realtime}. Starlink currently does not expose handoff control to users.

Common handoff strategies include selecting satellites with either maximum visible time or maximum RSS. The former reduces handoff frequency but may sacrifice signal quality; the latter improves signal quality but increases handoff overhead. Hence, we integrate handoff decisions into the QoE model and optimize satellite selection jointly with bitrate decisions.

\subsection{Throughput Sharing in LEO Network}
\label{ssec:motivate-multi-user}

\begin{figure}[!t]
    \centering
    \includegraphics[width=\columnwidth]{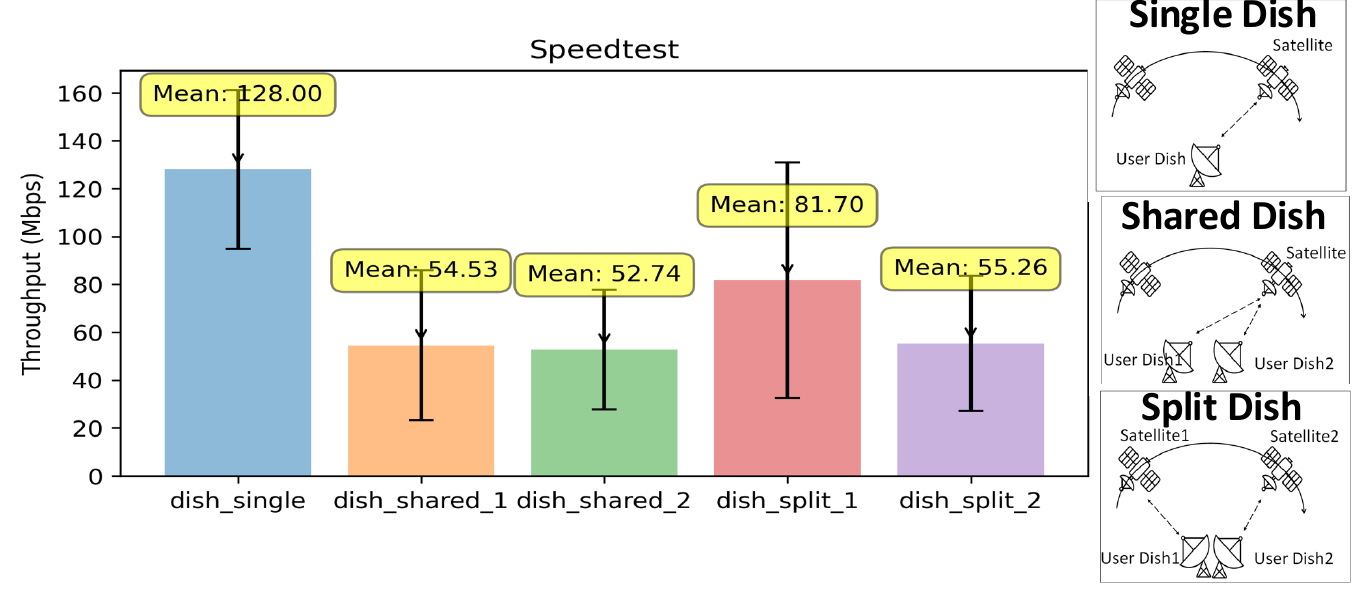}
    \caption{Throughput with (i) one dish, (ii) two dishes in the same direction, (iii) two dishes in different directions.}
    \label{fig:motivation_sharing}
\end{figure}

Throughput sharing among users significantly impacts QoE. Fig.~\ref{fig:motivation_sharing} shows measurements under three settings using Starlink: a single dish achieves $120$~Mbps; two co-located dishes sharing the same satellite reduce throughput to $50$~Mbps each; but directing the dishes to connect to different satellites improves aggregate throughput.

These results highlight the need for coordinated satellite selection in multi-user environments. We explore both distributed and centralized strategies for joint satellite and bitrate selection to optimize network-wide video QoE.

\section{Our Approach}
\label{sec:approach}

We formulate the joint optimization of video bitrate adaptation and satellite selection in $\S$\ref{ssec:problem}. Then, we describe our joint optimization for a single user in $\S$\ref{ssec:single-user} and joint optimization for multiple users in $\S$\ref{ssec:multi_user}. 

\subsection{Joint Optimization Problem Formulation}
\label{ssec:problem}

To support video streaming in LEO networks, one option is to adopt the same strategy as on the Internet, which decouples the video bitrate selection and satellite selection. While simple, this strategy leads to sub-optimal performance since satellite selection significantly impacts video streaming QoE. Therefore, we propose having clients perform joint satellite and video bitrate selection. 

Moving the handoff decision from the satellite to the end host is consistent with the trend in cellular networks, which have also moved from network-based handoff to client-based handoff, to not only reduce latency but also offload base stations. A LEO client can determine when to handoff and which satellite to handoff to because the satellites' trajectories are known in advance according to \cite{starlink.sx}. Moreover, the handoff can be performed efficiently on a per-packet basis by applying the corresponding beamforming weight to the received signals across its antenna array and passing the combined signal to the decoder for further processing. As the client knows its performance objective, having it select the satellite and video rate improves performance.

We define the QoE metric as follows to capture the impact of the satellite selection:
\begin{align}
\mathrm{QoE}_{\rm SAT}
=
&\sum_{k=1}^{N}
\bigl[\mu_{1} Q(R_{k}) - \mu_{2} T(R_{k}, \mathrm{Sat}_{k})\bigr]\nonumber\\
&-
\mu_{3} \sum_{k=1}^{N-1}
\left| Q(R_{k+1}) - Q(R_{k}) \right|
\label{eq:QoE-SAT}
\end{align}




Our joint ABR algorithm tries to optimize the QoE in Eq.\ref{eq:QoE-SAT}. The only difference from existing separate ABR algorithms is that ${\rm Sat}_k$ is considered an optimization variable. 

\subsection{Single-user QoE Optimization Algorithms}
\label{ssec:single-user}
In this section, we present two methods, MPC and RL, in LEO networks and analyze their respective characteristics.

\subsubsection{Joint MPC Algorithm}
\label{ssec:mpc}
\begin{algorithm}[t]
\caption{Joint pruned MPC}\label{alg:mpc}
\begin{algorithmic}[1]
\Require $Sat$: satellite, $t$: timestamp, $k$: video chunk index, $C, \hat{C}$: throughput, $Q$: QoE reward, $R$: bitrate, $B$: buffer size, $Sat_c$: the current satellite, $H$: handoff point
\State $\hat{C}_{Sat_c, t_k} = ThroughputPred(C_{Sat_c, [t_{k-N}, t_k]})$
\State $Q_{best}, R_{best} = f_{mpc}(R_{k-1}, B_{k-1}, \hat{C}_{Sat_c, t_k})$
\For {$Sat_n$ in $VisibleSatellites(t_k)$}
\State $\hat{C}_{Sat_n, t_k} = ThroughputPred(C_{Sat_n, [t_{k-N}, t_k]})$
\For {$H = 0$ to $F-1$} \Comment{$F$: future prediction length}
\State $Q, R = f^{sat}_{mpc}(R_{k-1}, B_{k-1}, H, \hat{C}_{Sat_c, t_k}, \hat{C}_{Sat_n, t_k})$
\If {$Q_{best} < Q$}
\State $Q_{best} = Q$
\State $R_{best} = R$
\State $Sat_{best} = Sat_n$
\State $H_{best} = H$
\EndIf
\EndFor
\EndFor
\State \Return $Q_{best}, R_{best}, Sat_{best}, H_{best}$

\end{algorithmic}
\end{algorithm}

MPC is effective for video bitrate adaptation on the Internet~\cite{favor, yin2015control}. It uses predicted throughput to optimize the QoE for the next few video chunks. It finds the best QoE by calculating all possible combinations of video quality for several future video chunks. Therefore, its performance highly depends on the quality of throughput prediction. To accommodate throughput prediction errors, there are several variants of MPC. RobustMPC improves the robustness of prediction against error by estimating prediction errors and using $C = \frac{Predict}{1 + \max\{error\}}$ as input. We adopt RobustMPC in our problem due to its resilience against errors. 


We propose two satellite-specific MPCs: joint exhaustive MPC and joint pruned MPC. Joint exhaustive MPC considers all combinations of satellites and bitrates for the future chunks download, which is the same logic as traditional MPC but additionally includes the satellites in the combination. However, this approach is computationally heavy due to its significant resource requirements. Its time complexity for a single calculation is $O(|R|^F \times |Sat|^F)$ where $|R|$ is the number of data rates, $|Sat|$ is the number of satellites, and $F$ is the number of future chunks considered. This cost is too high to run in real time. Therefore, we propose a joint pruned MPC assuming only one handoff during the next $F$ video chunks in the optimization horizon. This is a reasonable assumption since $F$ is generally small (\eg, $F=5$ in existing work) and multiple handoffs in such a short time window are expensive.
Joint pruned MPC significantly reduces computational overhead over joint exhaustive MPC. 
Its time complexity is $O(|R|^F \times |Sat| \times F)$ where $F$ represents the possible handoff point. We use joint pruned MPC in evaluation.


Algorithm~\ref{alg:mpc} shows a joint pruned MPC that considers both bitrates and satellites. The video player starts downloading a video chunk $k$ at $t_k$. We then predict the future throughput of the connected satellite \(S_c\) using \(ThroughputPred()\). We employ two prediction methods and compare their performances: (i) Harmonic mean over historical throughput; (ii) A model similar to \cite{vasisht2021l2d2} which takes elevation angle, azimuth angle, and weather information as input, and leverages machine learning models to predict future throughput. With the predicted throughput, \(f_{mpc}()\) applies the MPC algorithm. This algorithm tries to find the best QoE by iterating over all satellite candidates \(VisibleSatellites()\) and handoff points \(H\) when downloading future chunks. At each handoff point $H$, we use 
\(f^{sat}_{mpc}()\) to compute the utility by taking into account the handoff delay and throughput change when switching from \(Sat_c\) to \(Sat_n\) at \(t_{k+H}\).

\begin{figure}[t]
    \centering
    \includegraphics[width=1\columnwidth]{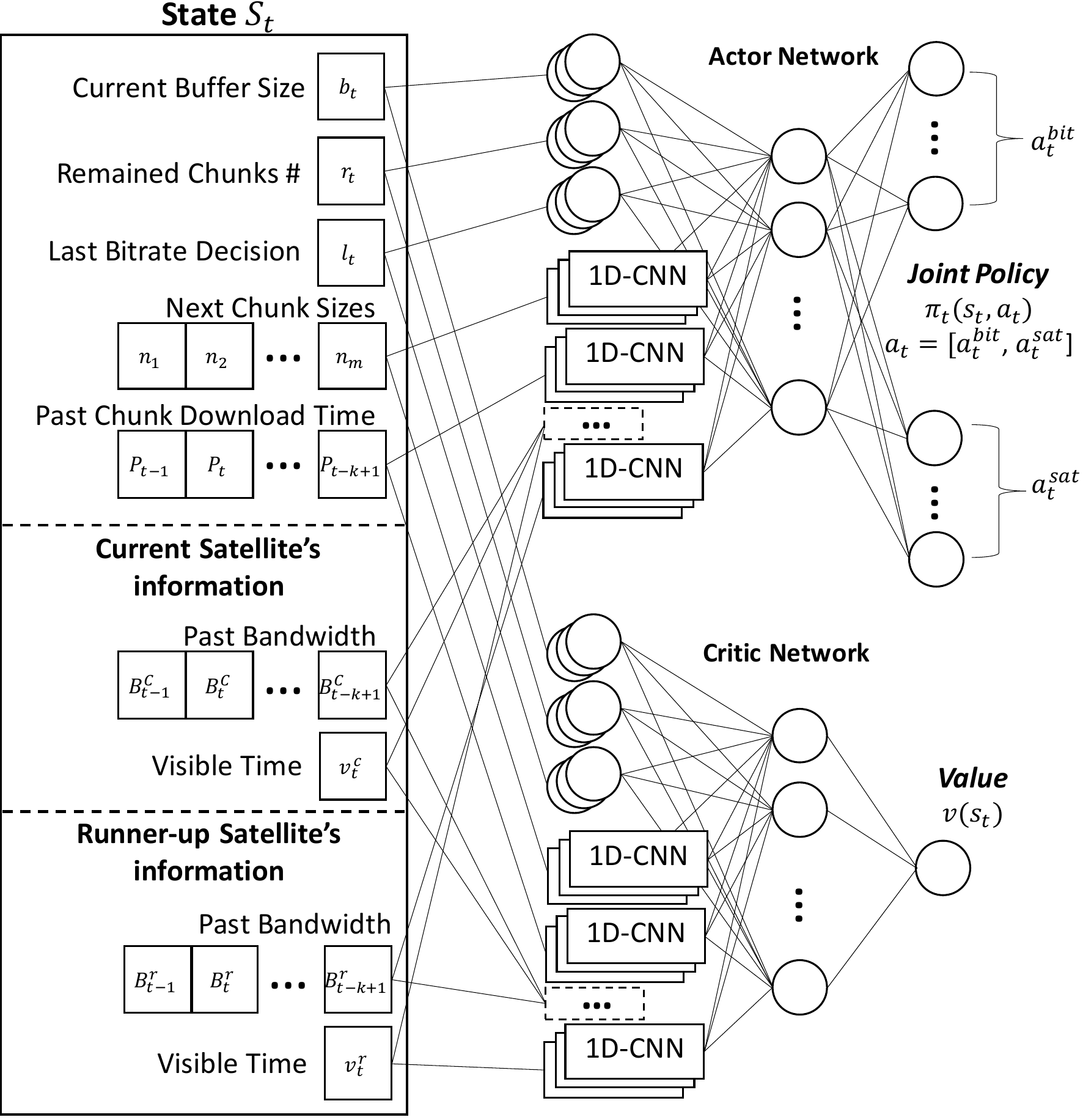}
    \caption{PPO-based RL generates joint optimization policies.}
    \label{fig:RL network}
\end{figure}

\subsubsection{RL-based Algorithm}
\label{sec:RL}

Our optimization problem is well-suited to an RL framework, where an agent learns to select the satellite and video bitrate that maximizes QoE. We use proximal policy optimization (PPO)~\cite{schulman2017proximal}, an actor-critic method that trains a critic to estimate the value function and an actor to optimize the policy while constraining policy updates for stability. Fig.~\ref{fig:RL network} illustrates the network with state space \(S\), value function \(v\), action space \(a\), and state transition probability \(\pi\).

\noindent \textbf{State:} At time \(t\), state \(S_t = (b_t, \gamma_t, l_t, n_m, P_t, B_t, v_t)\) includes current buffer level \(b_t\), remaining chunks \(\gamma_t\), last chunk bitrate \(l_t\), available next chunk sizes \(n_m\), download times \(P_t\), throughput measurements \(B_t\) for two satellites, and visible time \(v_t\). We extend prior work~\cite{mao2017neural} by adding \(B_t\) and \(v_t\) for both current and runner-up satellites, essential for joint bitrate and handoff decisions. Only past states are input to predict actions implicitly without separate predictors.

\noindent \textbf{Action:} Action \(a_t = [a^{bit}_t, a^{sat}_t]\) selects bitrate and satellite.

\noindent \textbf{Reward:} The video QoE (Eq.\ref{eq:QoE-SAT}) is the reward to optimize.

\noindent \textbf{Policy:} Policy \(\pi_t(s_t,a_t)\) gives probabilities of choosing bitrate and satellite from visible candidates. For efficiency, we limit candidates to two satellites (best and runner-up) in evaluation.

\noindent \textbf{Policy update:} The RL agent learns to maximize QoE by receiving rewards \(r\) per chunk. The critic estimates gradients guiding actor updates via the PPO policy gradient.

\noindent \textbf{Training process:} The algorithm collects user states and satellite throughput data, calculates individual QoE, tunes hyperparameters (learning rates, weights), recalculates QoE to avoid degrading other users, makes decisions based on updated QoE, and repeats this cycle until the session ends or user state changes significantly.

\subsection{Multi-user QoE Optimization Algorithms}
\label{ssec:multi_user}

In $\S$\ref{ssec:single-user}, we finally introduce several optimization algorithms for a single-user scenario. In practice, multiple user terminals can share the same satellite and compete for limited network resources. As shown in $\S$\ref{ssec:motivate-multi-user}, the throughput of each user significantly decreases when sharing with others. 

There are several ways to address this issue: (i) Let users independently make their own decisions using the single-user algorithm described in $\S$\ref{ssec:single-user}. The throughput is affected by cross traffic, making this approach sub-optimal and potentially unstable, as users optimize their own utility rather than the global one. (ii) Use a centralized algorithm that considers information from all users and makes decisions that maximize global utility; however, it faces deployment issues in practice, as it is challenging to control all users' selections of satellites and video rates. Motivated by the limitation of the above options, (iii) we design the RL that uses other users' states as input to optimize the global objective but only takes an action for a specific user at a time. This improves global coordination while remaining feasible to deploy, since it avoids controlling all users simultaneously. Last but not least, this RL design involves collecting other users' information in real time, which may still be difficult in some deployments. Therefore, (iv) we introduce \textit{centralized training and distributed inference} RL design. This approach uses all users' states as the input during training but only uses the current user's state during inference, which not only simplifies the deployment but also allows us to explicitly consider the interaction between multiple satellites and ground stations.

Our multi-user algorithms primarily aim to optimize the QoE for users engaged in video streaming. However, in real-world scenarios, LEO networks support a diverse range of applications beyond video streaming, including online gaming and web browsing. To better reflect these practical environments, we incorporate multiple users with dynamic resource allocation mechanisms, simulating the diverse network demands of different applications. This approach ensures a more realistic and adaptable optimization framework.

\comment{
\subsubsection{Multi-user MPC}

We extend the single-user MPC to a multi-user MPC as follows: (i) we gather all users' state information (\eg, XXX), and (ii) we optimize the global objective: the sum of all users' utilities. Note that since every user makes decisions at different times, during each invocation of MPC, we are allowed to only change one user's satellite and video bitrate selection and consider the impact of that user's selection on the global objective. 
}

\subsubsection{Multi-user RL (Joint RL (L))}
We apply the single-user PPO-based RL to the multi-user scenario. Our multi-user process can be represented by the following four tuples: ($S^j_t$, $A^j$, $\pi^j$, $R^j$), where $j$ is the user id, $S^j_t$ represents the user $j$'s state, $A$ is the distributed action space, $\pi$ is the state transition probability, and $R$ is the local reward value. The decision-making process of a single-user RL can be directly applied to multi-user optimization. This RL uses only the current user's state, so we call it the local state. It uses the throughput history of the current user's satellite to implicitly learn the network condition when sharing its satellite with other users. 

For multi-user training and testing, we use a critic network for distributed training. The input state of the critic network is a distributed environment (\ie, $S^j_t$ for each user). The critic network estimates the value of action decisions generated by the actor network. The state transition probability $\pi$ is then used to update the subsequent actor steps, evaluate each actor, and guide their parameter updates. The parameters and gradients of both networks are shared by all agents. 

\subsubsection{Multi-user RL with Global Status (Joint RL (G))}
This approach leverages all users' states to maximize all users' overall QoE, hence called Joint RL with global information (Joint RL (G)). This RL algorithm works by substituting the input states in the multi-user RL design with the following: ($S_t$, $A^j$, $\pi^j$, $R^j$) for each user $j$. The only difference is that $S_t$ represents all users' states, not just one user. We feed all users' information to the critic network to train a single global policy. 
During training, the centralized critic network is used to update the actor networks of all users. The gradient of the critic network is computed based on the actions taken by all users, and updates in the actor network are based on gradients in the critic network. Since it requires global information not only for training but also for the inference phase, it requires users to frequently exchange their states with other users who may share the same satellites. 


\begin{figure}[t]
    \centering
    \includegraphics[width=0.5\textwidth]{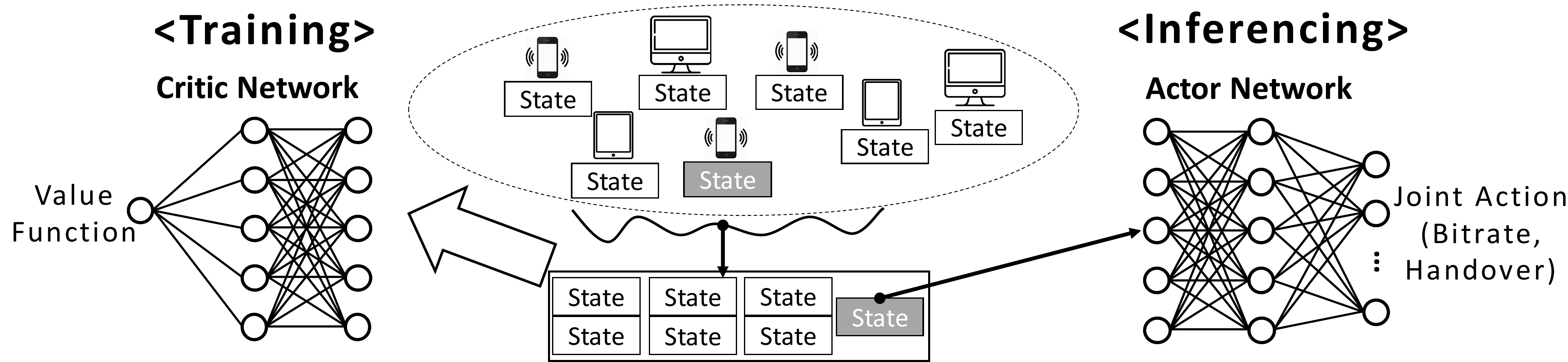}
    \caption{PPO-based RL algorithm generates joint optimization policies with centralized training and distributed inference.}
    \label{fig:dist_inference_cent_training network}
\end{figure}

\subsubsection{Multi-user RL with Centralized Training and Distributed Inference (Joint RL (L+))}
Inspired by the asymmetric actor-critic structure from the RL community \cite{pinto2017asymmetric,yu2022surprising,schneider2022deepcomp}, we use a centralized critic architecture and a decentralized actor architecture, 
as shown in Fig.~\ref{fig:dist_inference_cent_training network}. 
During training, the centralized critic network takes the global state as input and outputs a single value for each state, which is used to update the actor networks of all users.  This speeds up convergence and improves performance. 
During inference, the decentralized actor network takes the local state of each user as input and outputs a local action for that user. 
In other words, \textit{centralized training and distributed inference} have a hybrid concept of considering the global objective during training and making decisions at the user-level state during inference.


\comment{
\lili{not controlling all users: \lc{In summary, centralized RL on multiple-user communication involves a central controller that takes into account the status of all users and makes a decision that maximizes the overall QoE. This approach can lead to better performance compared to distributed approaches, where each user makes independent decisions without considering the impact on other users.}}
}

\subsubsection{Centralized MPC} 
So far, we have described several RL algorithms that can run in practice. We also consider the centralized MPC algorithm as another baseline. This algorithm assumes we have full control of all users' actions. While this assumption is strong and may not hold in practice, this is an interesting baseline. Without prediction error and with a large enough optimization horizon, the oracle centralized MPC establishes the upper bound for multi-user video QoE. We extend the single-user MPC to a centralized MPC by (i) treating all users' selected satellites and video bitrates as the optimization variables and (ii) taking the sum of all users' utilities as the objective. Since MPC calculates all the possible combinations, the search space increases exponentially with the number of users. To reduce running time, we prune the satellites with low throughput.

\comment{
\begin{small}
\begin{equation}
\max_{x_i} \sum_i QoE(x_i) \quad 
s.t. \sum_j x_j \leq 1
\label{eq:ratio-based}
\end{equation}
\end{small}
}

\section{System Implementation}
\label{sec:system-implementation}

To deploy our proposed approach in real-world scenarios, we address a key practical challenge: the absence of a native interface in Starlink that allows clients to actively select or switch between satellites. To overcome this limitation, we develop a custom client-server architecture that enables both bitrate selection and satellite handoff at the client side.

We emulate the satellite network using TCP sockets to represent distinct links between users and satellites. Upon a client’s connection to a satellite, a dedicated TCP socket is created on the server side and managed by a satellite controller. This controller enforces bandwidth constraints by rate-limiting the socket according to throughput traces collected from real-world Starlink deployments or from synthetic traces generated for experimentation. \zhiyuan{Note that the Starlink traces represent end-to-end throughput, encompassing both uplink and downlink, as they are collected from the connection between the video server and end users. When replayed in our system, they accurately emulate real-world conditions.}

When a client initiates a satellite handoff, the system closes the current socket and establishes a new one under a different satellite controller. Multiple clients may share a satellite, in which case their throughput is fairly allocated by the shared controller. To support multi-agent reinforcement learning (Joint RL(G)), clients report their internal states—such as buffer occupancy and download speed—to the server after each video chunk download. In response, the server returns the latest states of all other clients via a separate API call, enabling informed joint decision-making.


\noindent \textbf{RL implementation.} 
Our reinforcement learning model takes as input a sequence of past network measurements. These include download times and throughput values for the last $8$ chunks, which are processed through densely connected neural network layers with 128 neurons and 1D convolutional layers with 128 filters (kernel size 1, stride 1). These design choices are guided by a microbenchmark study.

The actor network uses a softmax activation at the output to generate a probabilistic policy over actions, while the critic network outputs a scalar value via a linear layer to estimate state values. Both networks use a learning rate of \(10^{-4}\), and we set the discount factor $\gamma$ to 0.99 to balance short-term and long-term performance. To control the exploration-exploitation trade-off, we incorporate an adaptive entropy weight that decays with training progression.

\noindent \textbf{Multi-user scenario.} 
To evaluate the system under realistic conditions, we simulate a network environment with 20 concurrent users performing video streaming or other background tasks. For users not engaged in video streaming, we introduce dynamic resource allocation to emulate heterogeneous and fluctuating network usage patterns.

To scale our reinforcement learning framework to multi-user settings, we adopt an encoder-decoder architecture within the neural network. This design allows the model to capture inter-user interactions efficiently and produce joint action decisions. While our current implementation targets scenarios with 20 users, the framework is extensible to larger populations by proportionally increasing model capacity.
\section{Evaluation}

We evaluate our proposed MPC and RL algorithms through extensive experiments in both single-user and multi-user scenarios, using simulation and testbed setups described in $\S$\ref{sec:system-implementation}. The video used is in 1080p, segmented into 49 chunks of 2 seconds each, and encoded at three bitrates: 300, 1200, and 2850~kbps. While real-world video servers typically offer six or more bitrates, we use three levels to manage the complexity of centralized MPC. To demonstrate scalability, we additionally evaluate \textbf{Joint RL (L+)} with six bitrates: 300, 750, 1200, 1850, 2850, and 4300~kbps. In the QoE function, the weights \(\mu_1\), \(\mu_2\), and \(\mu_3\) are set to 1, 4.3, and 1, respectively. The link RTT is 80~ms, and the handoff delay is 200~ms.

\subsection{Satellite Network Datasets}

We create three datasets: simulated, NOAA, and Starlink, each split into training and test sets. The simulated dataset models Starlink phase I satellite trajectories over 24 major cities worldwide using a free space path loss model to simulate throughput dynamics. The NOAA dataset contains measured SNR data from NOAA satellites, while the Starlink dataset comprises end-to-end measured throughput data from a real Starlink link between a client and a video server.

For MPC evaluation, algorithms are directly applied to test sets. For RL methods, models are first trained on the simulated dataset’s training set, then fine-tuned on the training set of NOAA or Starlink prior to testing.

The simulated throughput $b_t$ is generated as
$b_t = \alpha b_{\max} \frac{d_{\min}^2}{d_t^2} + \epsilon$,
where $b_{\max}$ is the maximum throughput of a Starlink satellite (Mbps),
$d_{\min}$ is the satellite altitude,
$d_t$ is the distance to the client,
$\alpha$ is a scaling factor to simulate resource limits,
and $\epsilon \sim \mathcal{N}(-2, 1)$ is Gaussian noise.
We generate 24 traces per city lasting 10 minutes each;
seven cities form the test set, and the rest are used for training.

Due to difficulty obtaining SNR from Starlink devices, we use NOAA satellites (NOAA-15, NOAA-18, NOAA-19) for SNR data. These LEO satellites have altitudes (808-854 km) and velocities (7.4 km/s) comparable to Starlink’s (550 km altitude, 7.6 km/s velocity), making NOAA traces suitable for handoff evaluation. We collect SNR and throughput traces from major US and Chinese cities, mapping elevation angle to SNR. Each Starlink satellite pass is assigned an SNR from a randomly chosen NOAA trace using this mapping, from which throughput is derived. The NOAA dataset contains 24 training and 12 test traces with no overlap.

The Starlink dataset is collected via iperf downloads from a Starlink RV terminal to a video cloud server in the US, capturing uplink, downlink, and terrestrial links. Logs are sampled every second over 13 hours. To simulate multiple satellites, we apply time-division multiplexing by converting a 16-minute log into four 4-minute logs, generating 48 multiplexed traces from the 13-hour log. Forty logs form the training set, and eight form the test set.

\subsection{Baseline Methods}

Our methods can select satellites and bitrates simultaneously. We compare them with popular satellite selection strategies and ABR algorithms. The following satellite handoff strategies are tested: 
\begin{itemize}
    \item Maximum Visible Time (MVT): Switch to the satellite with the longest visible time when the currently connected satellite is about to leave the client's horizon.
    \item Maximum Received Signal Strength (MRSS): Switch to the satellite with the highest signal strength.
    \item Maximum Bandwidth (MB): Switch to the satellite with the maximum available bandwidth when the currently connected satellite is about to leave the client's horizon.
\end{itemize}

\begin{figure}[t]
    \centering
    \includegraphics[width=\columnwidth]{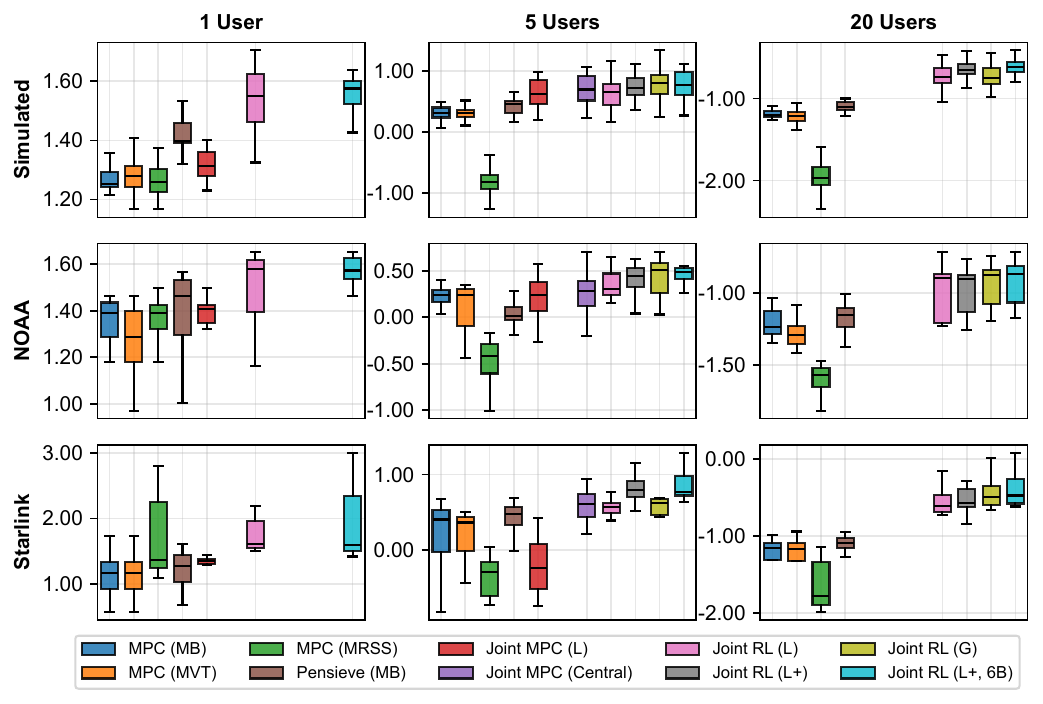}
    \caption{QoE results of our models (i.e., MPC, RL). The results are from simulated, NOAA, and Starlink datasets. All models use three bitrate levels for fair comparison, except Joint RL (L+, 6B). The number of users refers to how many users are streaming video out of 20 users.}
    \vspace{-10pt}
    \label{fig:simulation-overall}
\end{figure}
\begin{figure}[t]
    \centering
    \includegraphics[width=\columnwidth]{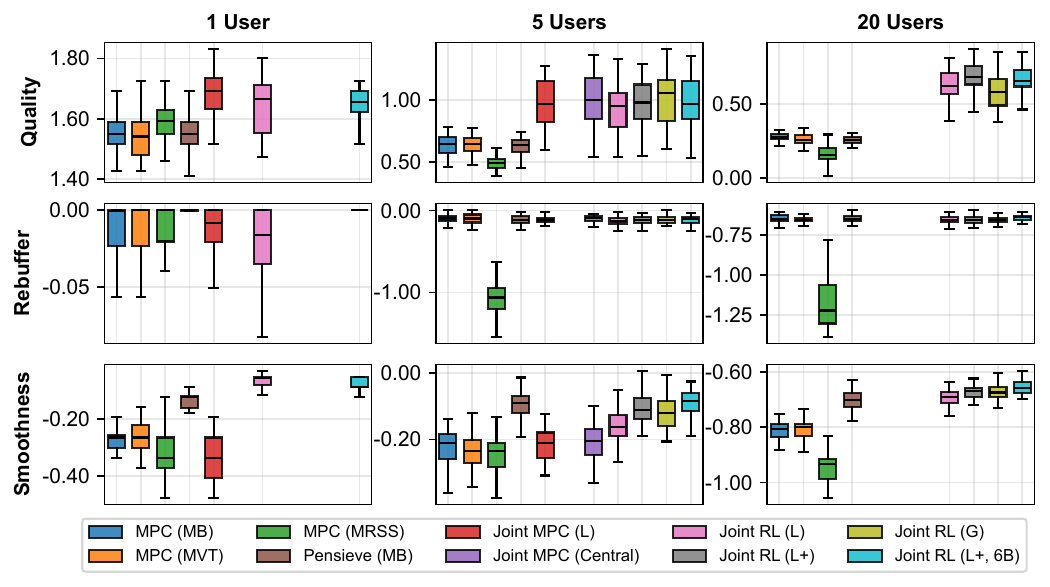}
    \caption{QoE breakdown with (1) quality reward, (2) rebuffer penalty, and (3) smoothness penalty. The results are from the simulated dataset. All models use 3 bitrate levels for fair comparison except Joint RL (L+, 6B). The number of users refers to how many users are streaming video out of 20 users.}
    \label{fig:simulation-reward-breakdown-simulate}
\end{figure}

\subsection{Simulation Results}

RobustMPC \cite{yin2015control} and Pensieve \cite{mao2017neural} are used in ABR algorithms. For simplicity, we refer to RobustMPC as "MPC". The satellite handoff strategy and the ABR algorithm are combined and tested. For example, MPC (MRSS) uses MRSS to select satellites and RobustMPC to decide bitrates. We pair Pensieve with the MB  due to its superior performance. We highlight \textbf{Joint RL (L+)} in bold to indicate that it is our final proposed model. We experiment with three scenarios involving 20 users: (i) 1 user streaming video while 19 users engage in other applications, (ii) 5 users streaming video while 15 users engage in other applications, and (iii) all 20 users streaming video.



\subsubsection{Overall Comparison}

The simulation results on three different datasets with different user numbers are presented in Fig.~\ref{fig:simulation-overall}. The proposed joint methods consistently outperform separate methods, particularly when multiple users are present. Compared to the best result achieved by the separate selection, joint selection methods can improve QoE by 18\%, 68\%, and 57\% for 1 user, 5 users, and 20 users, respectively, in the Starlink dataset. The substantial growth in QoE improvement with more users stems from our network resource allocation: while a single user enjoys ample resources, significant resource contention arises when three or more users share the network. In the simulated dataset, we train in 17 cities and test in 7 other cities. Our RL methods generalize well to unseen cities. Overall, the joint RL (G) has the best performance. Joint RL (L) and \textbf{Joint RL (L+)} are also competitive but slightly worse than Joint RL (G). Note that \textbf{Joint RL (L+)} always yields a better QoE than Joint RL (L), ranging from 0.8\% to 25\%. \kj{It shows that our \textit{centralized training and distributed inference approach} effectively considers both sides of the satellite network: periodicity of satellite movement and irregular network throughput.}

We present the breakdown of QoE into three terms (quality, rebuffer, and smoothness) in Fig.~\ref{fig:simulation-reward-breakdown-simulate}. Based on Fig.~\ref{fig:simulation-reward-breakdown-simulate}, separate selection methods generally produce lower-quality rewards when compared to joint selection methods. This can be attributed to a mismatch between the satellite selection strategy and the video QoE. Joint selection methods can actively change to a new satellite, which potentially brings QoE improvement. MB is the best handoff strategy to maximize the throughput among separate selection schemes, but it is still worse than the joint selection in quality score.

It is observed that MPC-based methods are often less smooth than RL-based ones, possibly due to the limited optimization window of MPC. This limitation also exists in our proposed Joint MPC method. Thus, in the 1-user scenario, although Joint MPC (L) produces higher quality rewards than Pensieve (MB), its final QoE score falls behind that of Pensieve (MB) due to worse smoothness. Due to the limited calculation window in MPC, we can see that the performance of Joint MPC (Central), which knows the network status within the window, is similar to or even underperforms our Joint RL.

Pensieve (MB) rarely picks the highest bitrate, even though it has a sufficient buffer. Hence, it has lower quality scores than our proposed Joint RL, as shown in Fig.~\ref{fig:simulation-reward-breakdown-simulate}. We interpret Pensieve (MB) as making a conservative decision, as it does not jointly consider satellites. When Pensieve (MB) increases the video quality due to enough buffers, it encounters many cases where a rebuffering penalty occurs due to sudden drops in satellite throughput.


We find that Joint RL actively decides the handoff to achieve a high QoE, \ie, 2 to 3 times more handoffs than the MB strategy. Unlike Pensieve (MB), Joint RL consistently selects the highest bitrate when the buffer is sufficiently full. Interestingly, Joint RL does the handoff only when the buffer is enough, which indicates a low possibility of rebuffering. Conversely, when the buffer is sufficient, Joint RL actively performs actions to achieve high throughput, such as performing multiple handoffs within a few seconds. 

By incorporating global status, Joint RL (G) achieves better results than Joint RL (L). Joint RL (G) learns interesting policies maximizing the overall QoE of all users, even if it slightly degrades one user. Joint RL (G) learns to tactically establish each user's role, such as determining handoff frequency or video quality. For example, one user does a lot of handoffs when encountering resource sharing and only maintains modest quality downloads. This allows other users to download the highest quality without resource contention. Impressively, our RL models learn these policies to optimize the overall QoE without any special a priori knowledge.

Compared to the simulation and Starlink trace for 5 users, the MPC models have long tails in the boxplot, indicating huge variances in QoE. This is because the real trace contains unstable latencies at specific timestamps, i.e., handoff. To cope with this unexpected case, users changed the quality abruptly as shown in the smoothness in Fig~\ref{fig:simulation-reward-breakdown-simulate}.  In contrast, the RL variants, including our proposed Joint RL (L+), yield highly stable QoE results, indicating that they can handle harsh network conditions more robustly. Note that a sudden one-time quality change incurs a severe degradation of QoE.



\begin{figure}[t!]
    \centering
    \includegraphics[width=\columnwidth]{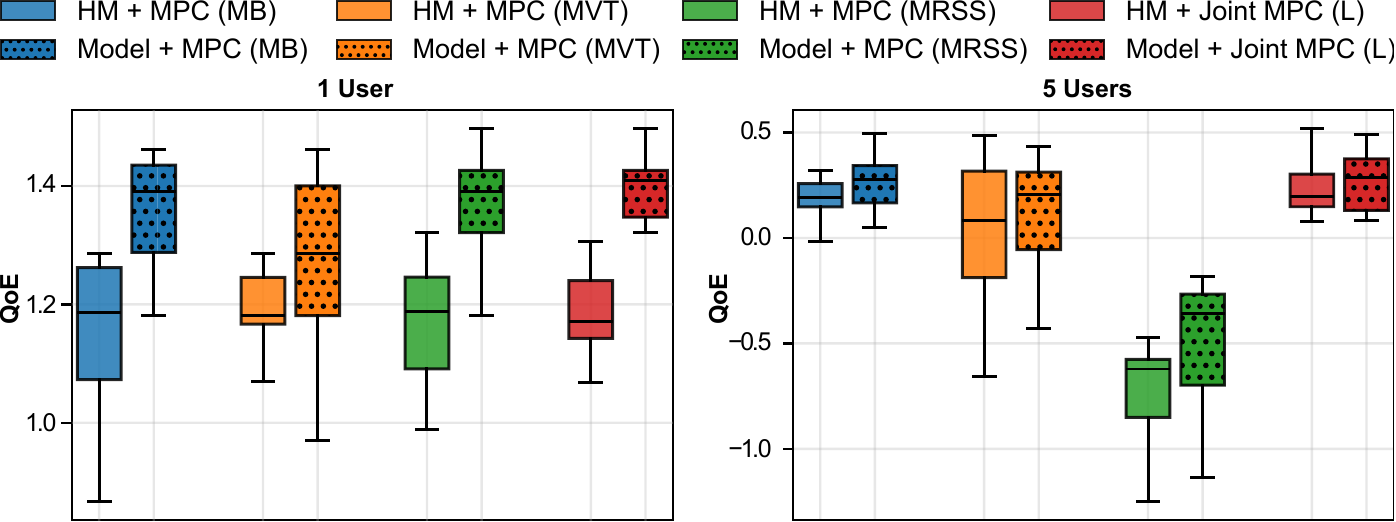}
    \caption{QoE results of MPC methods on the NOAA dataset with and without the model prediction. "HM" indicates the harmonic mean, while "Model" uses ML for prediction.}
    \vspace{-10pt}
    \label{fig:simulation-effect-of-model}
\end{figure}

\subsubsection{Obstructions}
\begin{figure}[t!]
    \centering
    \includegraphics[width=\columnwidth]{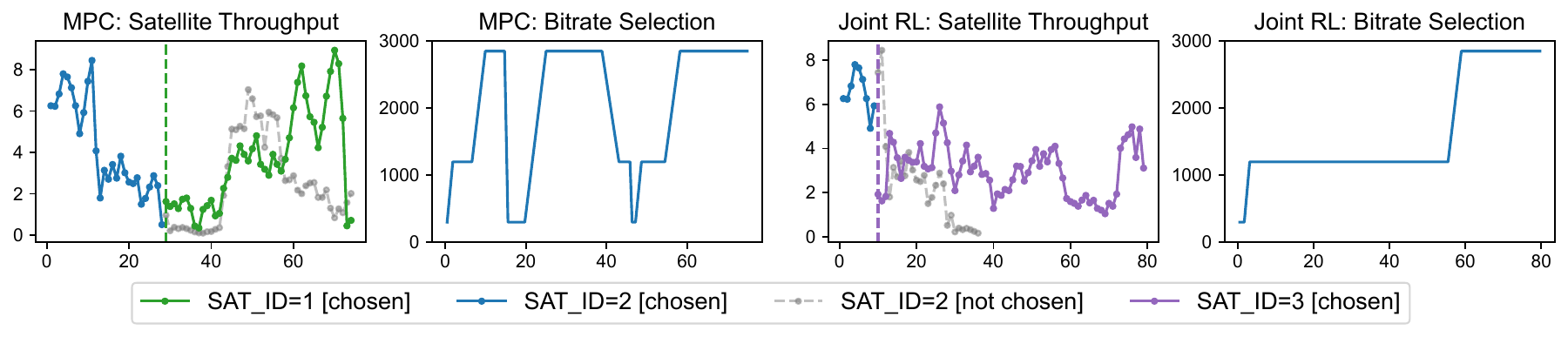}
    \caption{An obstruction scenario, with the X-axis denoting time (sec). Initially, both Joint MPC (L) and \textbf{Joint RL (L+)} connect with satellite 2. However, satellite 2 experiences an obstruction period from 15s to 40s. Joint MPC (L) switches to a different satellite around 30s with significant bitrate fluctuation. Conversely, \textbf{Joint RL (L+)} transits to an alternative satellite when disturbed, effectively maintaining stable bitrates.}
    \label{fig:obstruction-visualization}
\end{figure}

In Sec~\ref{ssec:motivate-video}, we highlight that obstructions decrease video QoE significantly in a satellite network. We find that our proposed methods can effectively handle the obstructed scenario. In the Starlink dataset, sometimes bandwidth drops abruptly, degrading QoE as shown in Fig.~\ref{fig:motivation_video_QoE_stanard}. If the bandwidth falls below the 25$th$ percentile for over 10 seconds, it is considered an "obstruction period". In all, 158 obstruction periods were identified in the dataset, accounting for 5.8\% of the total duration. Our Joint RL effectively handles these scenarios as shown in Fig.~\ref{fig:simulation-overall}, and we provide a case study in Fig.~\ref{fig:obstruction-visualization}. Unlike other methods that have bitrate fluctuations when faced with obstructions, RL methods smoothly switch to a different satellite and maintain consistently high bitrates.

\subsubsection{Effects of throughput prediction}
We employ an ML-based model similar to \cite{vasisht2021l2d2} to predict the throughput of the satellite using the NOAA dataset. The input features for the model are the azimuth angle, elevation angle, weather status, and the satellite's movement. The model uses a stacked ensemble of gradient-boosting decision trees and NN models to predict the future throughput, achieving a mean absolute error of $0.1304~Mbps$ on the test set.

Our prediction model is useful for throughput-based ABR algorithms. We compare the QoE of MPC-style methods with and without our model's predictions to evaluate its efficiency. Specifically, we compare the performance of MPC methods that use our model's predictions with those that use the harmonic mean of historical throughput to select bitrates, considering several handoff strategies.



As shown in Fig.~\ref{fig:simulation-effect-of-model}, model prediction leads to improved QoE for both traditional MPC methods and our proposed joint MPC. The throughput prediction enables better bitrate selection for traditional MPC methods, while for our proposed methods, it enables not only better bitrate selection but also better satellite selection.



\begin{figure}[!t]
    \centering
    \begin{minipage}[t]{0.48\columnwidth}
        \centering
        \includegraphics[width=\linewidth]{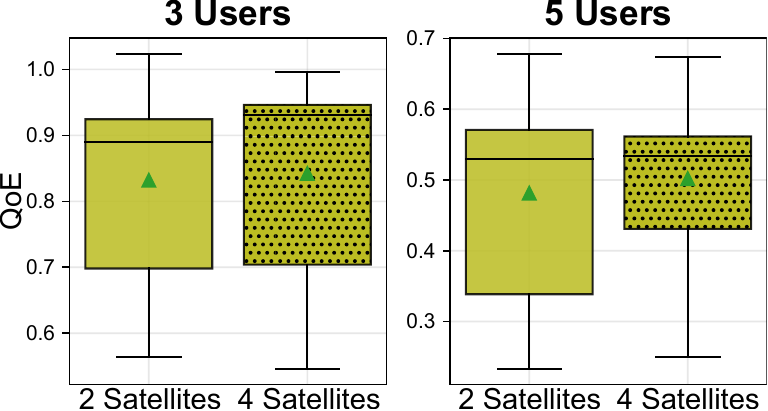}
        \caption{\textbf{Joint RL (L+)} on NOAA dataset varying satellite selection coverage.}
        \label{fig:multi_sats}
    \end{minipage}
    \hfill
    \begin{minipage}[t]{0.48\columnwidth}
        \centering
        \includegraphics[width=\linewidth]{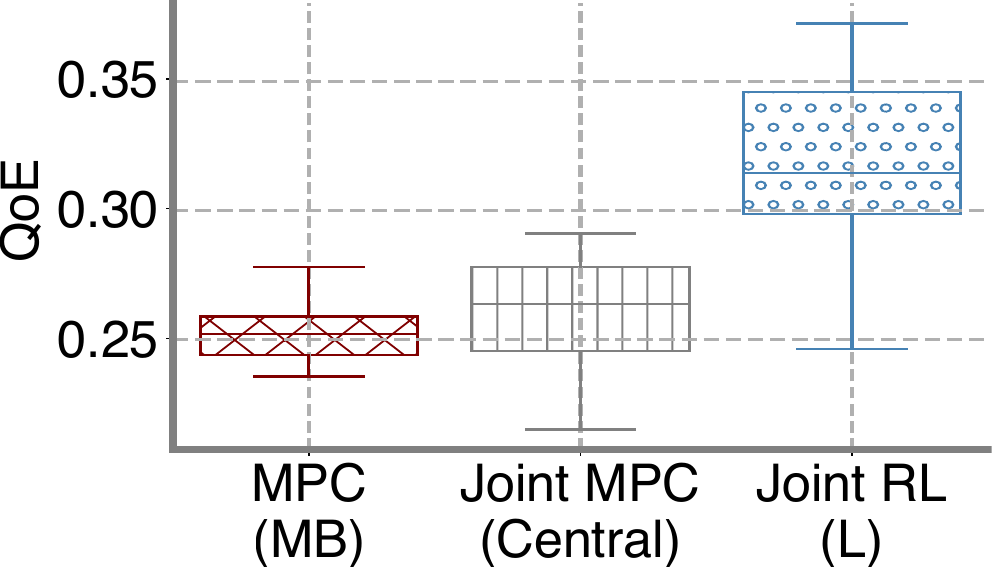}
        \caption{QoE results of three models for the multi-session problem (3 sessions).}
        \label{fig:multisession}
    \end{minipage}
\end{figure}

\begin{figure*}[!t]
    \centering
\includegraphics[width=1\textwidth]{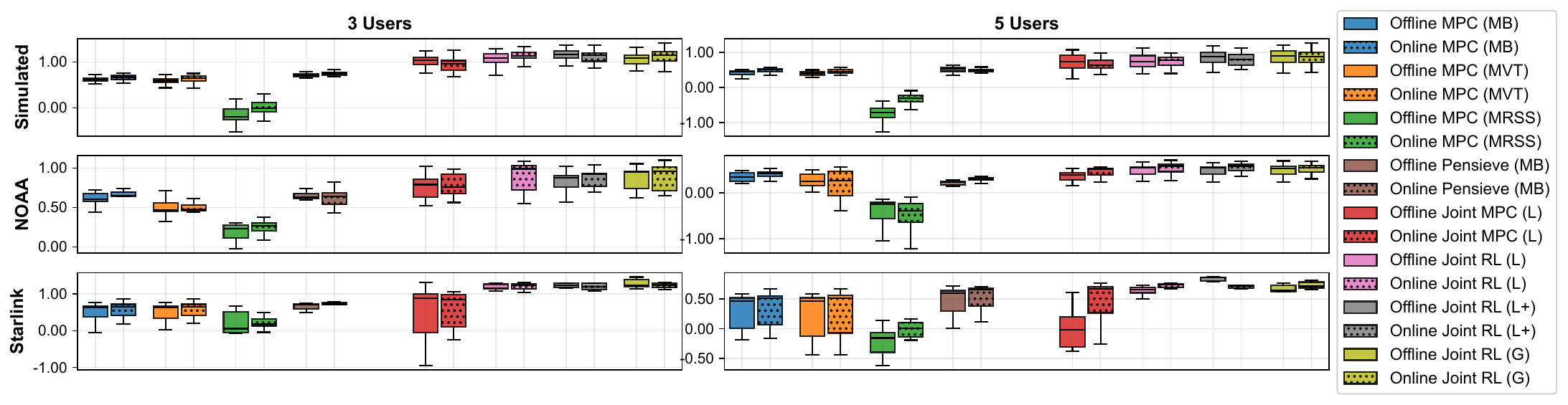}
\vspace{-10pt}
    \caption{Comparison of QoE results between testbed and simulation. Testbed results are masked by dots.}
     \vspace{-10pt}
    \label{fig:testbed-simulation-alignment}
\end{figure*}

\subsubsection{Consideration of multiple satellites}
Our models consider 2 satellites in the decisions. Fig.~\ref{fig:multi_sats} shows how much the QoE can be improved by considering more than 2 satellites. Since there are up to 4 visible satellites in NOAA, we evaluate the \textbf{Joint RL (L+)} by extending the input/output states to account for 4 satellites. The QoE of the model considering all satellites improved by 1.3\% and 4.1\% on average for 3 users and 5 users, respectively. Our RL design is scalable to support multiple satellites without major changes.

\begin{table}[t!]
\centering

\caption{QoE results for Joint MPC (Central) with two resource sharing strategies}
\label{tab:ratio_res}
\resizebox{0.3\textwidth}{!}{%
\begin{tabular}{ccc}
\hline
\textbf{Sharing Strategy} & \textbf{3 Users} & \textbf{5 Users} \\ \hline
Fair Resource             & 0.999            & 0.641            \\
Prioritized Ratio         & 1.032            & 0.761            \\ \hline
\end{tabular}
}
\end{table}

\subsubsection{Prioritized ratio strategy}
When total bandwidth is limited, the different resource-sharing strategies may matter. We experiment with two strategies: (i) fair resource sharing and (ii) prioritized ratio sharing on the simulated dataset using Joint MPC (Central). We use sequential least squares programming every second to find the optimized resource ratio of users. We filter only those tracks where more than two users are connected to the same satellite. We also apply a heuristic to calculate the rebuffering time by reducing the user's current buffer size by 70\%. This is to accommodate the unexpected events because even a small rebuffering time highly degrades QoE. When we tested using the original buffer sizes in the objective function, the algorithm fully utilized the buffered data; however, this is risky under prediction errors, which can generate rebuffering.

Table~\ref{tab:ratio_res} shows that the prioritized ratio strategy outperforms the fair strategy by about 3.3\% for 3 users and up to 18.7\% for 5 users. It indicates that a prioritized ratio strategy is beneficial when there is a lot of resource competition. Note that this prioritized strategy can only be applied in a centralized model because it should gather the states of all users, such as buffer sizes and previous bitrates, and make decisions for all users.

\subsubsection{Multi-session}
We evaluate the three models in multiple sessions, sharing network resources at a single user terminal. We assume that only one session can make a handoff decision to avoid unexpected issues (\eg, each session performing the handoff at different times). Since all sessions share the same user terminal, handoffs must be performed simultaneously. Fig.~\ref{fig:multisession} shows that Joint RL (L) outperforms MPC (MB) and Joint MPC (Central), which is in line with the other evaluations. This result indicates that our proposed models can work well in multi-session problems.

\subsubsection{Effects of locations}

\begin{table}[t!]
\centering
\caption{QoE comparison among cities (Joint RL (G))}
\resizebox{0.95\columnwidth}{!}{$
\begin{tabular}{@{}ccccccccl@{}}
\toprule
\textbf{City} & Hong Kong & Seoul & Beijing & Chicago & Toronto & Paris & London  \\ \midrule
\textbf{Latitude}     & 22.3 & 37.5 & 39.9 & 41.9 & 43.7 & 48.9 & 51.5 \\
\textbf{QoE - 3 Users}         & 0.87 & 1.04 & 0.96 & 1.10 & 1.08 & 1.16 & 1.31  \\
\textbf{QoE - 5 Users}       & 0.50 & 0.71 & 0.75 & 0.90 & 0.94 & 1.08 & 1.16  \\ \bottomrule
\end{tabular}
$}

\label{tab:city-comp}
\end{table}

According to the data presented in Table~\ref{tab:city-comp}, cities at higher latitudes tend to have higher video QoEs. This is likely because LEO satellites are more concentrated in high latitudes, whereas there are fewer satellites in low latitudes  \cite{mcdowell2020low}. These results come from a simulation. In the real world, other factors such as ground station distribution, obstruction, and user density can also affect the QoE.

\vspace{-5pt}
\subsection{Testbed Results}

To validate the effectiveness of our algorithms in practical scenarios, we implement them in a testbed system as described in $\S$\ref{sec:system-implementation}. As shown in Fig.~\ref{fig:testbed-simulation-alignment}, the results of the online (in testbed) and offline (in simulation) are consistent in most cases. It indicates that joint selection techniques can enhance the video QoE, and Joint RL can outperform other methods.

The major difference we observed in our testbed system is that HTTP requests may get stuck due to network fluctuation and packet loss in an online environment. To address this issue, we have set a timeout threshold of 10 seconds to download each chunk and immediately stop it on the client side if a request exceeds this limit, switch to another satellite, and attempt downloading again. This setting can improve performance, especially on Starlink datasets where obstacles can cause connection timeouts, as shown in Fig.~\ref{fig:testbed-simulation-alignment}. Besides, in the testbed, the latency becomes more unstable than in simulation, and the runtime of each algorithm can also affect performance. We found that our methods are efficient as the bitrate decision of each chunk takes less than $10~ms$. Thus, the runtime does not adversely impact system performance.

\section{Conclusion}

We propose a joint satellite selection and adaptive bitrate algorithm based on RL and MPC for LEO satellite networks. According to the experimental results, joint optimization can dynamically optimize satellite selection and bitrate decisions based on network conditions. Our results show that joint selection yields significant benefits in a single-user scenario, and this benefit further increases in a multi-user scenario. Future directions involve leveraging neural network-based video enhancement techniques at receivers to further optimize video streaming in LEO satellite networks (\eg, video super-resolution and interpolation upon packet losses).

\section{Acknowledgement}
This work is supported in part by NSF Grant CNS-2212297. The authors appreciate the insightful feedback from Yuqing Yang and IEEE INFOCOM 2026 anonymous reviewers.

\bibliographystyle{IEEEtran}
\bibliography{main}

\end{document}